\documentclass[icps3]{svjour}
\usepackage{graphicx}
\begin{document}

\title{Carrier relaxation with LO phonon decay in semiconductor quantum
   dots}

\titlerunning{Carrier relaxation with LO phonon decay in quantum dots}

\author{S.~A.~Levetas, 
M.~J.~Godfrey, P.~Dawson}

\authorrunning{S.~A.~Levetas, M.~J.~Godfrey, P.~Dawson}

\institute{Department of Physics, University of Manchester Institute of
Science and Technology, P.O.~Box~88, Manchester M60~1QD, United Kingdom,
e-mail: m.godfrey@umist.ac.uk}

\maketitle

\begin{abstract}
   Analysis of an exactly soluble model of phonons coupled to a
   carrier in a quantum dot provides a clear illustration of a phonon
   bottleneck to relaxation.  The introduction of three-phonon
   interactions leads to a broad window for relaxation by the
   processes of LO phonon scattering and decay.
\end{abstract}

\section{Introduction}
\label{sec:intro} 

The possible existence of a phonon ``bottleneck" that might inhibit
the relaxation of carriers in quantum dots remains a subject of
controversy.  Although the bottleneck is readily eliminated from
theoretical treatments by the introduction of non-phonon scattering
processes, there has also been interest in demonstrating relaxation by
intrinsic phonon scattering mechanisms alone.

Within the one-phonon approximation to the self-energy, the carrier's
spectral function (density of states) consists of a number of peaks
which broaden rapidly with increasing temperature.  This level
broadening is known not to be a lifetime effect \cite{InoshitaSakaki},
but might be supposed to help relieve the constraint of exact level
matching in LO phonon emission.  Indeed, the non-vanishing relaxation
rates found in simple non-equilibrium treatments
\cite{KralKhas,TsuchiyaMiyoshi} can be traced to the non-vanishing
spectral widths of the one-particle Green's function. 

\begin{figure}
  \includegraphics[width=\hsize]{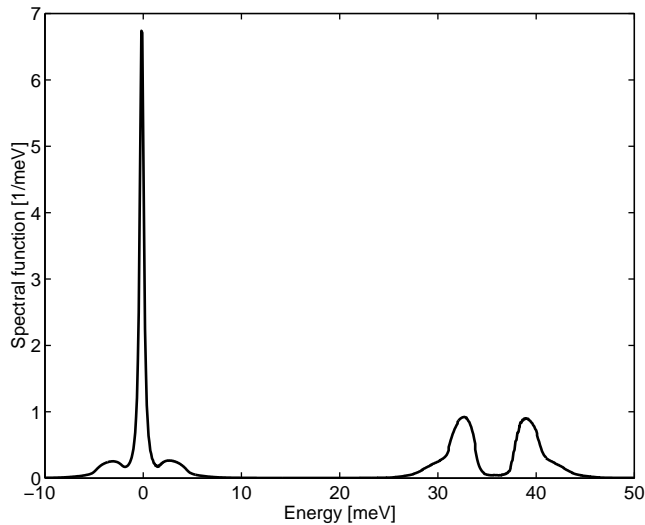} \caption{Electronic
  spectral function at $300\,$K for an unperturbed level separation
  $\Delta=\hbar\omega_0=36\,$meV.  The electron--phonon interaction
  is responsible for the Rabi splitting of the peak for the first excited
  state.} 
\label{fig:1}
\end{figure}

Though physically plausible, such theories conflict with the results
of experiments by predicting a vanishingly small relaxation rate in
the limit of low temperature and low excitation density, where Auger
processes can be neglected.  For this reason we have made a one-phonon
calculation of the carrier's spectral function in an exactly soluble
model for which we know that there can be \emph{no} relaxation, as
will be explained in Sect.~\ref{sec:solublemodel} below.
Figure~\ref{fig:1} illustrates the broadening of the spectral
peaks that---in this case, at least---is a spurious result due to the
one-phonon approximation.  Accordingly, one should regard with caution
any proposed relaxation mechanism that relies on the broadening
obtained in this widely-used approximation.

Despite the impossibility (in our model) of obtaining relaxation by
LO-phonon emission alone, we show in Sect.~\ref{sec:relaxation} that
the introduction of three-phonon interactions (leading, e.g., to the
decay of an LO phonon to give two LA phonons) can give a broad,
temperature-dependent window for relaxation, as was found previously
\cite{LiNakayamaArakawa} by use of the Wig\-ner--Weiss\-kopf method.

\section{A soluble model}
\label{sec:solublemodel}


Our model for phonons interacting with an electron in a parabolic
confining potential is the usual electron--phonon Hamiltonian with a
special choice of matrix elements $\sqrt{n+1}\,g_q$ between
neighbouring electron levels,
\begin{eqnarray}
  H
  &=&
  \Delta\sum_n n\,c^\dagger_n c_n + \sum_q \hbar\omega_q b^\dagger_q b_q
  \nonumber \\
  &+& \sum_{n,q} \sqrt{n+1}\,g_q\bigl(b_q+b^\dagger_{-q}\bigr)
  \bigl(c^\dagger_{n+1}c_n+c^\dagger_{n}c_{n+1}\bigr)\,, 
\label{eqn:elH}
\end{eqnarray}
where $c_n$ and $b_q$ are the electron and phonon annihilation
operators.  The calculation of the spectral function illustrated in
Fig.~\ref{fig:1} was based on Eq.~(\ref{eqn:elH}), truncated to four
levels.  In the one-electron subspace, the ladder operators
$a=\sum_n\sqrt{n+1}\, c^\dagger_nc_{n+1}$ and $a^\dagger$ obey Bose
commutation rules and the above Hamiltonian is equivalent to
\begin{eqnarray}
  H
  =
  \Delta a^\dagger a + \sum_q \bigl\{\hbar\omega_q b^\dagger_q b_q
  + g_q\bigl(b_q+b^\dagger_{-q}\bigr)\bigl(a+a^\dagger\bigr)\bigr\}\,.
\label{eqn:solubleH}
\end{eqnarray}
The Hamiltonian (\ref{eqn:solubleH}) represents a system of coupled
oscillators, so that all features of its energy spectrum and response
can be obtained without approximation.  For example, the response to
an impulsive electric field applied at time $t=0$ is given by the
retarded susceptibility
\begin{eqnarray}
  \chi(t)= -i\langle X(t)X(0)-X(0)X(t)\rangle\,\theta(t)\,,
\label{eqn:chidef}
\end{eqnarray}
where $X=a+a^\dagger$ is proportional to the electron dipole moment
and $\theta(t)$ is the unit step function.  In the frequency domain,
$\chi$ satisfies the Dyson equation
\begin{eqnarray}
  \hbox{$\chi(\omega) = \chi^0(\omega) 
  + \chi^0(\omega)\sum_q|g_q|^2D^0_q(\omega)\,\chi(\omega)$}\,,
\label{eqn:chiDyson}
\end{eqnarray}
which can be solved for $\chi(\omega)$ in terms of the
unperturbed phonon Green's functions~$D^0_q(\omega)$. 

For coupling to dispersionless LO phonons the calculation can be
completed analytically.  The spectral function
$-2\,\hbox{Im}\,\chi(\omega)$ consists of a pair of (Rabi split)
delta-function peaks that result from the binding of a phonon to the
electron~\cite{InoshitaSakaki}.  In the time domain, $\chi(t)$ is the
sum of two undamped oscillations: the electric field excites the electron
into a state which is a superposition of exact eigenstates from the
discrete spectrum of~(\ref{eqn:solubleH}), and from which the electron
is unable to relax.

For coupling to LA phonons, the sum appearing in (\ref{eqn:chiDyson})
can be evaluated numerically.  We find that if $\Delta$ falls in the
LA phonon continuum, relaxation is possible by emission of an LA
phonon, but the rate is many orders of magnitude too slow for single
LA-phonon emission to be significant as a mechanism for relaxation.
%

\section{Influence of three-phonon interactions}
\label{sec:relaxation}

Finally we turn to mechanisms that can lead to the fast relaxation of
an electron in a quantum dot.  Interactions between phonons of
branches $i$, $j$, and $k$ can be described by a phenomenological
Hamiltonian~\cite{Klemens}
\begin{eqnarray}
  H^{ijk} = \beta\!\!\!\sum_{p+q+r=0}
        \omega_p^i\, \omega_q^j \,\omega_r^k\,
        X_p^i\, X_q^j\, X_r^k\,,
\label{eqn:KlemensH}
\end{eqnarray}
where $X^i_p=\bigl(b^i_{p}+b_{-p}^{i\,\,\dagger}\bigr) /
\bigl(M^i\omega^i_p\bigr){}^{1/2}$.
We choose the coefficient $\beta$ (the same for all $ijk$) to obtain
agreement with the observed lifetime of free LO phonons with respect
to decay to two LA phonons, $\tau=2.5\,$ps in GaAs at $300\,$K.  Note
that we must, more generally, consider also the interactions with
$i=\hbox{LA}$ and $j=k=\hbox{LO}$, because an LO phonon bound to an
electron can be liberated by absorption or emission of an LA phonon
with simultaneous relaxation of the electron.  In
contrast, non-dispersive free LO phonons cannot be scattered in this
way because of the constraint of energy conservation.

The effect of the interactions is calculated perturbatively.  The
LO-phonon Green's function $D_q(\omega)$ satisfies a Dyson equation
\begin{eqnarray}
  D_q(\omega) = D^0_q(\omega)+D^0_q(\omega)\,\Pi_q(\omega)\,D_q(\omega)\,,
\label{eqn:phonDyson}
\end{eqnarray}
where the phonon self-energy $\Pi_q(\omega)$ is evaluated to second
order in~$\beta$ by use of the Matsubara technique~\cite{Mahan}.
%
%
The phonon Green's function obtained in this way replaces
$D^0_q(\omega)$ in Eq.~(\ref{eqn:chiDyson}), which gives an
approximation for the susceptibility for the electron.   A
relaxation rate for the electron is estimated by fitting an
exponential \cite{LiNakayamaArakawa} to the time-domain
susceptibility,~$\chi(t)$.

\begin{figure}
  \includegraphics[width=\hsize]{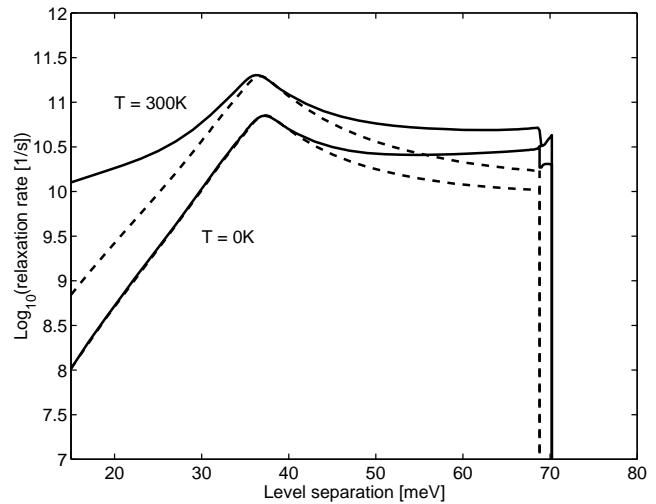}
  \caption{Carrier relaxation rate as a function of level separation.
  Dashed lines are results for the mechanism
  $\hbox{LO}\rightarrow2\,\hbox{LA}\,$;  
  results shown by solid lines also include
  $\hbox{LO}\rightarrow\hbox{LO}\pm\hbox{LA}$.}   
  \label{fig:2}
\end{figure}

Results of these calculations are shown in Fig.~\ref{fig:2}, for
temperatures $T=0$ and $300\,$K, and for a range of electron level
separations.  In all cases the relaxation rate is greatest when the
unperturbed level separation matches the LO phonon energy,
$\Delta\simeq\hbar\omega_0$.  This is understandable on qualitative
grounds, as in this region the electron--phonon bound-state wave
function contains nearly equal amplitudes for the electron to be in
its ground state (with one additional phonon) as in its first excited
state, so that the interaction Hamiltonian has a large matrix element
between the bound state and the electron ground state.  A new result
of our work is that the relaxation rate remains high only up to a
threshold value of $\Delta$ beyond which the energy conservation law
cannot be satisfied; in fact, as can be seen from the graph, there are
slightly different thresholds for the two kinds of relaxation process
considered.

The scattering process $\hbox{LO}\rightarrow\hbox{LO}-\hbox{LA}$ has a
particularly strong dependence on temperature.  Below resonance, it
leads to no enhancement to the relaxation rate at low temperature
because there are few LA phonons available to be absorbed.  In
contrast, at $300\,$K the absorption of LA phonons enhances the
relaxation rate by more than an order of magnitude at the smallest
values of~$\Delta$, and so provides the major route for relaxation in
this r\'egime.

\begin{acknowledgement}
\label{sec:thankyou}

The authors thank the EPSRC (UK) for support under grant no.\ GR/L81697.

\end{acknowledgement}

\end{document}